\documentclass[aps,twocolumn,superscriptaddress,prb]{revtex4-2}
\usepackage[english]{babel}

\usepackage{amsmath,graphicx,csquotes,cases,physics,mathtools,amssymb,siunitx,comment}
\usepackage[colorlinks=true, allcolors=black]{hyperref}
\usepackage{pgfplots}[compat=1.18]
\usepgfplotslibrary{fillbetween,polar}
\pgfplotsset{compat = newest}
\usepgfplotslibrary{external} 
\usepackage{caption, subcaption}
\newcommand{\bea}{\begin{eqnarray}}
\newcommand{\ea}{\end{eqnarray}}
\newcommand{\eea}{\end{eqnarray}}

\def\h{\hat}

\newcommand{\red}[1]{\textcolor{black}{#1}} 
\newcommand{\blue}[1]{\textcolor{black}{#1}} 

\begin{document}

\title{\red{Enhanced Electron Reflection} at Mott-Insulator Interfaces}

\author{Jan Verlage} 

\affiliation{Fakult\"at f\"ur Physik and CENIDE, Universit\"at Duisburg-Essen,
  Lotharstra{\ss}e 1, 47057 Duisburg, Germany,}




\author{Peter Kratzer} 
\affiliation{Fakult\"at f\"ur Physik and CENIDE, Universit\"at Duisburg-Essen,
  Lotharstra{\ss}e 1, 47057 Duisburg, Germany,}




\date{\today}

\begin{abstract}
The Klein paradox describes an incoming electron being scattered at a supercritical barrier to create electron-positron pairs, a phenomenon widely discussed in textbooks. 
While demonstrating this phenomenon experimentally with the fundamental particles remains challenging, condensed matter analogs are more accessible to experimental realization. 
For spinless quasi-particles, theoretical works show an enhancement of the pair production rate, and 
analogs of this effect in condensed matter systems have been studied theoretically. 
Here, we present another condensed matter system, a heterostructure comprised of two materials with 
strongly and weakly interacting electrons, 
that allows for constructing analytical solutions using the hierarchy-of-correlations method. 
\red{The results show enhanced electron reflection related with the production of doublon-holon pairs, as known from the Klein paradox.}
\end{abstract}

\maketitle

\section{Introduction}
The mathematical structure of physics sometimes facilitates striking conjectures by analogy between remote areas of research, e.g. between relativistic quantum theory and solid state physics. 
For instance, already 
in 1927 Oskar Klein \cite{Klein1929} investigated the solution of the Dirac equation for an incoming electron scattering (in one dimension) at a barrier of height $V$. 
His mathematical analysis indicated a non-vanishing transmission for barrier heights $V>2 m_e c^2$ exceeding twice the rest mass, even 
if the energy $E$ of the incoming particle 
is much lower than the potential height, $E< V$. Classically, this transmission is forbidden and
thus must be considered 
a purely quantum mechanical effect. 
It is called the \enquote{Klein paradox} \cite{Dombey1999,Hansen1981}. 
Subsequent theoretical studies by a large number of researchers 
\cite{Bjorken1964,Hansen1981,Greiner1985, Manogue1988, Holstein1998, Nitta1999, Merad2000, Dombey2000, Krekora2004, Krekora2005, Hejcik2008} 
attempted to shed light onto the apparent paradox, using 
single-particle quantum mechanical  or quantum field theoretical methods, 
without capturing all relevant aspects.
The former approach, asserting particle number conservation,  cannot correctly predict the pair creation associated with the supercritical barrier, while the latter does describe the pair creation correctly, but 
misses a proper treatment of the incoming electron. 
A comprehensive picture, including quantum field theory and 
the appropriate boundary condition for an 
incoming particle, required a numerical treatment  \cite{Krekora2004, Krekora2005}, and indicated a suppression of the pair creation rate by the incoming electron as it blocks available states because of the Pauli principle.
Analogous realizations of the Klein paradox for fermions led, for example, to perfect transmission in graphene \cite{he2013chiral,katsnelson2006chiral} and the solid-state realization in gapless two-dimensional materials was reported recently \cite{Gutierrez2016,Stander2009,katsnelson2006chiral}.

While all of these works investigate the Klein paradox in 
its original fermionic context, 
the bosonic analog is less well studied \cite{wagner2010bosonic,Mayoral2011,Brito2015}, and its space-time resolved analysis has received less attention \cite{Manogue1988,Brezin1970,Narozhny1974}. Theoretical space-time resolved works \cite{wagner2010bosonic} show that, in the absence of Pauli blocking, the incoming boson {\it enhances} the pair production rate, in a fashion similar to the stimulated emission of light from an excited atom \cite{wagner2010bosonic}. 
For experimental investigations with fundamental particles the high energy barrier needed \cite{Bjorken1964} turned out to be an obstacle. 
Additionally, a bosonic Klein paradox necessitates the existence of bosonic antiparticles, which usually signals instabilities~\cite{Harms2022}.
However, there are analogs to the bosonic Klein paradox ~\cite{Kleinherbers2025,Bassant2024,Harms2022} in condensed matter systems, one example being magnons at ferromagnetic interfaces.
Here, the anti-particle is a negative-energy magnon, and therefore there are no problems with instabilities. In these 
systems, the enhancement of the pair production  
has been used as an experimental signature of the bosonic Klein paradox. 

In this work, we present a theory for 
the interface of weakly and strongly correlated materials
and show that it provides an solid-state analog for studying the Klein paradox. 
The particles and anti-particles are the doublons and holons in a Mott insulator, such that there are no instabilities as in the magnon case. 

\red{The analogy to electron-positron pair creation in quantum electrodynamics at low energies has already been discussed in other settings like ultra-cold atoms in optical lattices \cite{Pineiro2019,Witthaut2011,Cirac2010,Zhu2007,Hou2009,Lim2008,Boada2011,Goldman2009,Kasper2016,Queisser2012,Szpak2012,Szpak2011}, electrons in semiconductors \cite{Linder2018,Smolyansky2009,Hrivnak1993} or $^3\mathrm{He}$ \cite{Schopohl1992}.
}
\blue{In advantage over the above systems, the analogy presented in this work comes even closer to the situation in quantum electrodynamics, as} the Mott gap in the Fermi-Hubbard model of the Mott insulator arises naturally through the interaction. Moreover, the quantitative analogy to the $\texttt{1+1}$ dimensional Dirac equation emerges without any fine-tuning. The particle-hole symmetry is analogous to the charge symmetry in the relativistic system. 
\red{The analogy between the doublon-holon pair creation and the electron-positron pair creation was already shown for specific Mott-Neel type spin backgrounds \cite{queisser2025doublon}. Intuitively, the similarity is between the lower Hubbard band and the Dirac Sea in quantum electrodynamics, as well as between the upper Hubbard band and the positive energy continuum \cite{lenarvcivc2012dielectric,oka2010dielectric,oka2005ground}.}
\red{However, one should keep in mind that the doublons and holons in the Mott insulator are neither bosons nor fermions, but composite particles.} 
\blue{Thus, it remained unclear up to now if the reflection at an interface can carry a signature of doublon-holon pair production in the Mott phase.}

This paper is organized as follows: we first introduce the hierarchy of correlations used to derive the propagation equations of the (quasi-)particles, 
followed by the scattering at the single interface, for which we derive the transmission and reflection coefficients. Lastly, we 
cast the problem into the form of 
a \red{Dirac} equation for the quasi-particles in the Mott insulator to 
\blue{complete the analogy with elementary particle physics.}

\section{Hubbard Model and Hierarchy of Correlations}
In order to describe both the weakly interacting layer as well as the strongly interacting Mott insulator, we employ the Hubbard model \cite{hubbard1963}, defined as follows:
\begin{equation}
\label{eq:FHMHamiltonian}
\hat{H} = -\frac{1}{Z}\sum_{\mu \nu s} T_{\mu \nu} c_{\mu s}^\dagger c_{\nu s} + \sum_\mu U_\mu \hat{n}_{\mu \uparrow} \hat{n}_{\mu \downarrow} + \sum_{\mu s} V_\mu \hat{n}_{\mu s}.
\end{equation}
Here, \(c_{\mu s}^\dagger\) and \(c_{\nu s}\) denote fermionic creation and annihilation operators acting on sites \(\mu\) and \(\nu\), respectively, with spin index \(s \in \{\uparrow, \downarrow\}\). The corresponding number operators are given by \(\hat{n}_{\mu s} = c_{\mu s}^\dagger c_{\mu s}\). The lattice structure and associated hopping amplitudes are encoded in the adjacency matrix \(T_{\mu \nu}\), which is taken to be nonzero only for nearest-neighbor pairs, where it assumes the value \(T\); otherwise, \(T_{\mu \nu} = 0\). The coordination number \(Z\) counts the number of nearest neighbors per site.

\red{The parameter $U_\mu$ describes the Coulomb interacting, and is only non-zero in the Mott insulator.} It is worth noting that in weakly correlated semiconductors, a small on-site interaction \(U_\mu\) may be treated within a mean-field framework and effectively absorbed into a renormalized \(V_\mu\), akin to the treatment in Fermi liquid theory \cite{landau1957theory,solovyev2017renormalized}.

\red{The on-site potential is, in principle, non-zero in both the strongly and weakly interacting layer, as the Mott bands are not necessarily centered around $U/2$, but might be shifted up- or downwards. Without loss of generality, as only the band alignment is relevant, we take $V_{\mu \in \mathrm{Mott}}\equiv 0$ and deal with the band alignment only by $V_{\mu \in \mathrm{semi}}$.} 

Therefore, the parameters \(U_\mu\) and \(V_\mu\) distinguish between strongly correlated systems \((U_\mu \neq 0, V_\mu = 0)\) and weakly correlated systems \((U_\mu = 0, V_\mu \neq 0)\)

\subsection{Hierarchy of Correlations}
To approximate solutions for charge modes at the interface, we employ the hierarchy of correlations~\cite{Queisser2014,Queisser2019,PhysRevA.82.063603,ywd2-bh2w}, which is particularly well-suited for systems with a large coordination number \(Z \gg 1\). In this framework, reduced density matrices involving two or more lattice sites are decomposed into on-site and correlated components. Specifically, for a pair of lattice sites \(\mu\) and \(\nu\), this decomposition reads \(\hat{\rho}_{\mu \nu} = \hat{\rho}_\mu \hat{\rho}_\nu + \hat{\rho}_{\mu \nu}^{\text{corr}}\). Based on the assumption \(Z \gg 1\), one can perform an expansion in powers of \(1/Z\), leading to a natural suppression of higher-order correlators.
The two-point correlation scales as \(\hat{\rho}_{\mu \nu}^{\text{corr}} = \mathcal{O}(Z^{-1})\), while the three-point correlation satisfies \(\hat{\rho}_{\mu \nu \lambda}^{\text{corr}} = \mathcal{O}(Z^{-2})\), and so on. 

\red{Performing this expansion into the inverse coordination number, the evolution equations form an infinite hierarchy:}
\begin{eqnarray}
    i \partial_t \hat{\rho}_\mu &=& F_1(\hat{\rho}_\mu, \hat{\rho}^{\text{corr}}_{\mu \nu}), \nonumber \\
    i \partial_t \hat{\rho}^{\text{corr}}_{\mu \nu} &=& F_2(\hat{\rho}_\mu, \hat{\rho}^{\text{corr}}_{\mu \nu}, \hat{\rho}^{\text{corr}}_{\mu \nu \lambda}).
\end{eqnarray}
\red{The exact form of the functionals $F_n$ depends on the specific structure of the Hamiltonian. By using the fact that \(\hat{\rho}_{\mu \nu}^{\text{corr}} = \mathcal{O}(Z^{-1})\) we approximate this to:}
\begin{equation}
    \begin{aligned}
        i \partial_t \hat{\rho}_\mu &\approx F_1(\hat{\rho}_\mu, 0), \quad \text{with solution } \hat{\rho}_\mu^{0}, \\
        i \partial_t \hat{\rho}^{\text{corr}}_{\mu \nu} &\approx F_2(\hat{\rho}_\mu^{0}, \hat{\rho}^{\text{corr}}_{\mu \nu}, 0).
    \end{aligned}
\end{equation}
This hierarchy yields an iterative scheme for approximating the full density operator \(\hat{\rho}\). Further details of the method are provided in Appendix~\ref{app:hierarchyofcorrelations}.

In analogy with the Hubbard \(X\) operators~\cite{Hubbard1965,ovchinnikov2004hubbard} and composite operator methods~\cite{mancini2004hubbard}, we introduce quasi-particle operators defined as:
\begin{equation}
\hat{c}_{\mu s I} = \hat{c}_{\mu s} \hat{n}_{\mu \bar{s}}^I = 
\begin{cases}
\hat{c}_{\mu s}(1 - \hat{n}_{\mu \bar{s}}), & \text{for } I = 0, \\
\hat{c}_{\mu s} \hat{n}_{\mu \bar{s}}, & \text{for } I = 1,
\end{cases}
\end{equation}
where \(I=1\) corresponds to doublons and \(I=0\) to holons. \red{These are the physical excitations within the Mott insulator on top of the half-filled background.} 
From these operators, we define the corresponding correlation functions \(\langle \hat{c}_{\mu s I} \hat{c}_{\nu s J} \rangle^{\text{corr}}\)~\cite{verlage2024quasi}. 

\subsection{Factorization}
For the relevant dynamics, these correlators may be factorized~\cite{navez2014quasi,Queisser2014} as:
\begin{equation}
\langle \hat{c}_{\mu s I} \hat{c}_{\nu s J} \rangle^{\text{corr}} = (p_{\mu s}^I)^* p_{\nu s}^J,
\end{equation}
where \(p_{\mu s}^1\) and \(p_{\mu s}^0\) are the doublon and holon amplitudes, respectively. These amplitudes can be combined into a spinor representation, from which the equations of motion for the (quasi-)particles can be derived \cite{verlage2024quasi}. \red{In a sense, this is the same as factorizing the many-body density operator $\h \rho$ in this doublon-holon basis as $\hat{\rho}_{IJ}=\left(p^I \right)^* p^J$ with the wave functions $p^I$.}

\red{In the interface system, the translational invariance is broken in only one direction, in the other ones it is still intact. Therefore, we can decompose the quasi-particle wave functions into their parallel momentum \(k^{\parallel}\) dependent Fourier components. After this, the index $\mu$ is just a scalar counting the layers parallel to the interface. In a hyper-cubic lattice the Schrödinger-like equations for doublons $I=1$ and holons $I=0$ can be combined using  \(U_\mu^I = I U_\mu\):}
\begin{equation}
	\label{difference}
\begin{aligned}
	\left(E-U_\mu^I-V_\mu\right)p_{\mu s}^I + \langle\hat{n}_{\mu \bar{s}}^I\rangle^0\sum_J T_{\bf k}^\| p_{\mu s}^J \\
	= -T\frac{\langle\hat{n}_{\mu \bar{s}}^I\rangle^0}{Z}\sum_J \left(p_{\mu-1 \,s}^J+p_{\mu+1\, s}^J\right).
\end{aligned}
\end{equation}
 The expectation values \(\langle \hat{n}_\mu^I \rangle^0\) are evaluated with respect to the mean-field background state \(\hat{\rho}^0_\mu\).
 \red{$T_{\bf k}^\|=2T/Z \sum_i \cos(k_i^\|)$ is the kinetic energy contribution from the bands formed in the translational invariant directions for nearest neighbor hopping.} For hyper-cubic lattice this means $Z=4$ in two dimensions and $Z=6$ in three dimensions. Moreover, this choice is not just guided by mathematical simplicity, but also by pervoskite structures in which the essential Mott physics happens on a cubic lattice and heterostructures from these material class can be grown with high precision.
A detailed derivation of this equation is provided in Appendix~\ref{app:hierarchyforFermiHubbard} or \cite{verlage2024quasi}. For more details about interface systems within the hierarchy see, for example, \cite{ywd2-bh2w,verlage2024quasi}.

\red{To first order, the two spin sectors decouple. Because of this, we will work from now on in the spin $\uparrow$ sector and drop the index $s$.}
\blue{In our expansion,} \red{the dynamics of the spin fluctuations is suppressed by an additional order $1/Z$, such that we treat the charge modes dynamic as it happens on top of a fixed spin background. }


\subsection{Mott-Neel Type Spin Order}
\red{The starting point of our analysis is the mean-field background. It is needed to analyze the effective equation \ref{difference}. The Mott insulator state has one particle per site, it is half-filled. This leaves the spin degree of freedom. In this work, we consider} the antiferromagnetic Mott-N\'eel state 
with its checkerboard structure:
\begin{equation}
\label{eq:rho0MN}
	\hat{\rho}_\mu^0= \begin{cases}
		\ket{\uparrow}\bra{\uparrow} & \mu \in A, \\
		\ket{\downarrow}\bra{\downarrow} & \mu \in B. \\
	\end{cases}
\end{equation}
\red{Because of the sublattice structure, an additional index $A$ and $B$ is necessary.} 
\red{The expectation values are fixed as $\langle \h n_{\mu_A \uparrow} \rangle=1$, $\langle \h n_{\mu_B \uparrow} \rangle=0$.}
The weakly interacting layer coupled to this will inherit the bi-partite structure. For this, we take either the valence band $\ket{\uparrow \downarrow}$ or the conduction band $\ket{0}$. Because our calculation is done at zero temperature, these two are related via the particle-hole symmetry.

\red{From the spin ordering of the background,} it follows directly that:
\begin{equation}
E\, p_\mu^{1_A} = (E - U)\, p_\mu^{0_B} \equiv 0.
\end{equation}
Hence, in the Mott-N\'eel state the coupled equations for doublons and holons thus read:
\begin{equation}
\label{eq:hierarchyMott}
	\begin{aligned}
		E p_\mu^{0_A}&= - \left[T_\mathbf{k}^\| p_\mu^{1_B}+\frac{T}{Z}\left( p_{\mu+1}^{1_B}+p_{\mu-1}^{1_B}\right)\right] ,\\
			\left(E-U\right) p_\mu^{1_B}&= - \left[T_\mathbf{k}^\| p_\mu^{0_A}+\frac{T}{Z}\left( p_{\mu+1}^{0_A}+p_{\mu-1}^{0_A}\right)\right].
	\end{aligned}
\end{equation}
In the weakly interacting layer, on the other hand, we find:
\begin{equation}
\label{eq:hierarchyweakly}
	\begin{aligned}
		(E-V) p_\mu^{0_A}&= - \left[T_\mathbf{k}^\| p_\mu^{0_B}+\frac{T}{Z}\left( p_{\mu+1}^{0_B}+p_{\mu-1}^{0_B}\right)\right] ,\\
		\left(E-V\right) p_\mu^{0_B}&= - \left[T_\mathbf{k}^\| p_\mu^{0_A}+\frac{T}{Z}\left( p_{\mu+1}^{0_A}+p_{\mu-1}^{0_A}\right)\right].
	\end{aligned}
\end{equation}

\section{Quasi-Particle Propagation}
In order to calculate the transmission characteristics at the interface between the weakly correlated layer and the strongly correlated Mott insulator, we first need to derive the propagation within the individual regions. Otherwise we could not use the correct \textit{ansatz} for the incoming, reflected and transmitted spinor.

\subsection{Strongly Correlated Layer}
From Eq.~\ref{eq:hierarchyMott} we can read off that there is a proportionality between the particle $p_\mu^{1_B}$ and hole $p_\mu^{0_A}$ solution on their respective sublattices, they are not independent. 
Thus, in the \textit{ansatz} $p_\mu^{1_B}=\mathcal{B}\kappa^\mu$ and 
$p_\mu^{0_A}=\mathcal{A}\kappa^\mu$
\red{the amplitudes $\mathcal{A}$ of the holons and $\mathcal{B}$ of the doublon wave functions are related.}  
\blue{Their relative sign allows us to distinguish two cases, }
$p_\mu^{1_B}=\beta p_\mu^{0_A}$ with $\beta= \pm \sqrt{\frac{E}{E-U}}$. 
\red{The one with the plus is called the even spinor, the other one the odd spinor. The odd one has a phase shift from one sublattice to the other.}

We find the eigenmodes:
\begin{equation}
\label{eq:eigenmodesMott}
	\begin{aligned}
		\kappa_{1,2}=&-\frac{Z}{2T} \left(\sqrt{E(E-U)}+T_\mathbf{k}^\| \right)\\&\pm\sqrt{\left[\frac{Z}{2T} \left(\sqrt{E(E-U)}+T_\mathbf{k}^\| \right) \right]^2-1},\\
		\kappa_{3,4}=&+\frac{Z}{2T} \left(\sqrt{E(E-U)}-T_\mathbf{k}^\| \right)\\&\pm\sqrt{\left[\frac{Z}{2T} \left(\sqrt{E(E-U)}-T_\mathbf{k}^\| \right) \right]^2-1}\\
	\end{aligned}
\end{equation}
with $\kappa_1 \kappa_2=\kappa_3 \kappa_4=1$. \red{Within the Hubbard bands, they are complex numbers obeying $|\kappa_i|=1$, such that $p_\mu^{1_B}$ and $p_\mu^{0_A}$ describe plane waves. This can be seen by applying the identity $x \pm i \sqrt{1-x^2}=e^{\pm i \arccos(x)}$, which defines the effective wave numbers for the propagation. They read:}
\begin{equation}
    \begin{aligned}
        \cos(x_{1,2})&=\frac{Z}{2T}\left(\sqrt{E(E-U)}+T_\mathbf{k}^\| \right) \\
        \cos(x_{3,4})&=\frac{Z}{2T}\left(\sqrt{E(E-U)}-T_\mathbf{k}^\| \right).
    \end{aligned}
\end{equation}
The proportionality between them reads:
\begin{equation}
	\label{eq:BiAieq}
	\mathcal{B}_i = \frac{1}{U-E}\left[T_\mathbf{k}^\|+\frac{T}{Z}\left(\kappa_i + \kappa_i^{-1}\right) \right]\mathcal{A}_i,
\end{equation}
which yields the $\beta$ with the correct sign. $\kappa_1$ and $\kappa_2$ \red{belong to the even solutions, i.e. having the same sign on neighboring sites of both sublattices, while} $\kappa_3$ and $\kappa_4$ \red{belong to the} odd solutions \red{with a sign change between sublattices}. The propagation direction of the individual modes is dictated by the group velocity: 
\begin{equation}
	v_G(\kappa_i)=-i \frac{1}{\kappa_i} \frac{\mathrm{d}\kappa_i}{\mathrm{d}E},
\end{equation}
which reads for the four modes
\begin{equation}
	\begin{aligned}
		v_G(\kappa_1)=-v_G(\kappa_2)&=\frac{Z}{8T}\frac{2E-U}{  \sqrt{E(E-U)}} \\& \frac{1}{\sqrt{1- \left[\frac{Z}{2T} \left(\sqrt{E(E-U)}+T_\mathbf{k}^\|  \right) \right]^2}},\\
		v_G(\kappa_3)=-v_G(\kappa_4)&=-\frac{Z}{8T}\frac{2E-U}{ \sqrt{E(E-U)}}\\&  \frac{1}{\sqrt{1- \left[\frac{Z}{2T} \left(\sqrt{E(E-U)}-T_\mathbf{k}^\|  \right) \right]^2}} \quad.
	\end{aligned}
\end{equation}
There is a sign change in the middle of the Mott gap $U/2$. For the lower energies, $\kappa_2$ and $\kappa_3$ describe right-propagating solutions, for higher ones $\kappa_1$ and $\kappa_4$ do this. This is relevant to find the correct \textit{ansatz} for the transmitted spinor.
Inside the Mott bands the group velocity is purely real, while in between the bands, $0 \leq E \leq U$, the group velocity turns complex. This describes both a propagation and a non-constant norm of the spinor, and already gives a first hint towards to Klein paradox analog. Outside the bands, the group velocity is purely imaginary, resulting in decaying solutions.

\begin{table}
		\caption{Group velocity of the Mott-N\'eel solutions inside the bands}
	\label{tab:groupvelMottin}
	\begin{center}
		\begin{tabular}{ c| c | c }
			& lower Hubbard band & upper Hubbard band \\ \hline
			$v_G(\kappa_1)$ & $<0$  & $>0$  \\  
			$v_G(\kappa_3)$ & $>0$ & $<0$     
		\end{tabular}
	\end{center}

\end{table}

\subsection{Weakly Correlated Layer}
In the semiconductor adjacent to the Mott insulator in 
the Mott-Ne\'el state 
we also impose the bi-partite lattice structure, such that we can again define the checkerboard sublattices $A$ and $B$. In this, the proportionality reads $p_\mu^{0_A}= \alpha p_\mu^{0_B}$. This $\alpha$ describes the symmetry of the hole solution $p_\mu^{0_X}$ between the two sublattices. There is an even symmetry with $p_\mu^{0_A}=  p_\mu^{0_B}$ and an odd one with $p_\mu^{0_A}= - p_\mu^{0_B}$. Both of these solve Eq.~\ref{eq:hierarchyweakly}.
Together with the \textit{ansatz} $\lambda^\mu$, we find the eigenmodes for $\alpha=+1$ as 
\begin{equation}
\label{semiconductor}
	 \lambda_\pm = -\frac{Z}{2T} \left(E-V+T_\mathbf{k}^\| \right) \pm \sqrt{\left[ \frac{Z}{2T} \left(E-V+T_\mathbf{k}^\| \right)\right]^2-1}.
\end{equation}
This is the even solution and has a group velocity 
\begin{equation}
\begin{aligned}
	v_G(\lambda_+)=&-i \frac{1}{\lambda_+}\frac{\mathrm{d}\lambda_+}{\mathrm{d}E}\\&=\frac{Z}{2T}\frac{1}{\sqrt{1-\left(\frac{Z}{2T}\left((E-V+T_\mathbf{k}^\|)\right)^2\right)}}>0.
\end{aligned}
\end{equation}
$\lambda_+$ describes right-propagating solutions, $\lambda_-$ left-propagating ones. \red{The effective wave number is a solution of $\cos(k)=\frac{Z}{2T}(E-V+T_\mathbf{k}^\|)$}.

Similar for $\alpha=-1$, we find with $p_\mu^{0_A}=\rho^\mu$:
\begin{equation}
	\rho_\pm = +\frac{Z}{2T} \left(E-V-T_\mathbf{k}^\| \right) \pm \sqrt{\left[ \frac{Z}{2T} \left(E-V-T_\mathbf{k}^\| \right)\right]^2-1}.
\end{equation}
This is the odd solution with 
\begin{equation}
\begin{aligned}
	v_G(\rho_+)&=-i \frac{1}{\rho+}\frac{\mathrm{d}\rho+}{\mathrm{d}E}\\&=-\frac{Z}{2T}\frac{1}{\sqrt{1-\left(\frac{Z}{2T}\left((E-V-T_\mathbf{k}^\|)\right)^2\right)}}<0.
	\end{aligned}
\end{equation}
For the odd solution $\rho_\pm$ the propagation direction is reversed. $\rho_+$ describes left-propagating solutions, $\rho_-$ right-propagating ones. These are the negative energy modes related to the Klein paradox analog. \red{The effective wave number is a solution of $\cos(k)=\frac{Z}{2T}(E-V-T_\mathbf{k}^\|)$}.

\begin{figure}
	\centering
	\begin{tikzpicture}
		\begin{axis}[
			axis lines=middle,
			xtick={0.001, 1.5708,3.1415}, 
			xticklabels={$0$, $\pi/2$,$\pi$}, 
			ytick={-0.4, 0.3, 1}, 
			yticklabels={$V-\frac{4T}{Z}$, $V$, $V+\frac{4T}{Z}$}, 
			ylabel={$E$},
			xlabel={$\kappa$},
			xmin=-0.1, xmax=3.4, 
			ymin=-0.45, ymax=1.2, 
			samples=100,
			domain=0:3.2,
		width=0.4\textwidth, height=0.3\textwidth,
		legend style={at={(1.,0.7)}, anchor=west}
			]
			\addlegendimage{empty legend}
			\addlegendentry{$k^\|$}
			\addplot[blue!100!red, smooth] {-0.04999999999999999 - 0.35*cos(deg(x))};
			\addlegendentry{$0\pi$};
			\addplot[blue!50!red, smooth] {0.0525126 - 0.35*cos(deg(x)) };
			\addlegendentry{$0.25\pi$};
			\addplot[blue!00!red, smooth] {0.3 - 0.35*cos(deg(x))};
			\addlegendentry{$0.5\pi$};

			\addplot[blue!100!red, dashed] {0.65  + 0.35*cos(deg(x))};
			\addplot[blue!50!red, dashed] {0.547487  + 0.35*cos(deg(x)) };
			\addplot[blue!00!red, dashed] {0.3 + 0.35*cos(deg(x))};
			
		\end{axis}
	\end{tikzpicture}
	\caption[Energy bands in bi-partite semiconductor]{Energy bands as a function of momentum $\kappa$ for different parallel momenta $k^\|$ for the even (solid lines) and odd (dashed lines) spinor solution in the weakly interacting layers.}
	\label{fig:bandersemioverlap}
\end{figure}
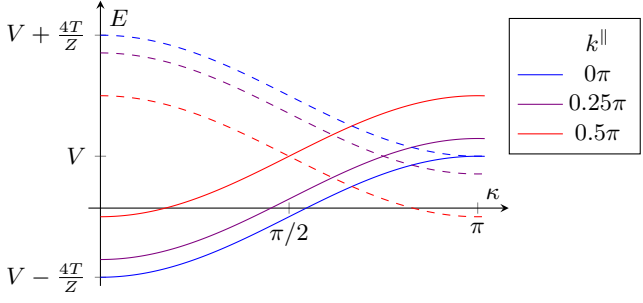

The dispersion relation $E(\kappa)$ in the weakly interacting region is shown in Fig.~\ref{fig:bandersemioverlap} for different parallel momenta. For any non-zero parallel momentum there is a band overlap at $E=V$, allowing scattering from one band into the other.

\subsection{Probability Current}
The effective (quasi-)particles in the weakly and strongly correlated half spaces are the electrons and doublons/holons, respectively. In order to physically make sense out of the transmission coefficients we need to introduce a probability current $J_\mu$ by the continuity equation:
\begin{equation}
	\partial_t D_\mu = J_\mu -J_{\mu-1}
\end{equation}
with the quasi-particle density:
\begin{equation}
	D_\mu = \sum_{\ell,X} |p_\mu^{\ell_X}|^2.
\end{equation}
From the general evolution for the bi-partite lattice case Eq.~\ref{difference}, we find the quasi-particle current as:
\begin{equation}
\label{eq:probCurrentEq}
	J_\mu = i \frac{T}{Z} \sum_{i l X} N_\mu^{iX} \left( p_{\mu+1}^{l \bar{X}} \left(p_\mu^{iX} \right)^*- p_{\mu}^{i X} \left(p_{\mu+1}^{l \bar{X}} \right)^*\right).
\end{equation}

\section{Scattering at the Interface}
We want to calculate the transmission coefficients at the interface for an incoming electron in the even spinor traveling from the weakly interacting half-space at $\mu \leq 0$ onto the interface with the Mott insulator at $\mu >0$. For this scenario, the \textit{ansatz} in the weakly interacting space reads 
\begin{equation}
	\label{eq:MNAnsatzSemi}
	\psi_{\mu \leq 0}=\begin{bmatrix}
		p_\mu^{0A} \\
		p_\mu^{1A} \\
		p_\mu^{0B} \\
		p_\mu^{1B} \\
	\end{bmatrix} = \frac{1}{2}\left(\lambda_+^\mu +\frac{R_1}{\lambda_+^\mu}\right)\begin{bmatrix}
		0 \\ 1\\0\\1
	\end{bmatrix}+\frac{1}{2}\frac{R_2}{\rho_-^\mu} \begin{bmatrix}
		0 \\ 1\\0\\-1
	\end{bmatrix}.
\end{equation}
	and in the Mott insulator 
	\begin{equation}
    \label{eq:MNAnsatzMott}
		\psi_{\mu>0}=  \frac{1}{\sqrt{2}} \frac{1}{N_i}\kappa_i^\mu \begin{bmatrix} \mathcal{A}_i \\ 0 \\0 \\ \mathcal{B}_i \end{bmatrix} + \frac{1}{\sqrt{2}} \frac{1}{N_j}\kappa_j^\mu \begin{bmatrix} \mathcal{A}_j \\ 0 \\0 \\ \mathcal{B}_j \end{bmatrix}.
	\end{equation}
The choice of $\kappa_i$ and $\kappa_j$ depends on the energy.
Inserting this \textit{ansatz} into Eq.~\ref{eq:probCurrentEq} quickly gives the incoming and reflected current:
\begin{equation}
	\begin{aligned}
		j^\mathrm{in}&=i \frac{T}{Z} \left[\lambda_+ - \lambda_+^{*}\right], \\
		j^\mathrm{ref}&=-i \frac{T}{Z}\left(|R_1|^2 \left[\lambda_+ - \lambda_+^{*}\right]- |R_2|^2 \left[\rho_- - \rho_-^{*}\right] \right).
	\end{aligned}
\end{equation}
The incoming current is only comprised, as per \textit{ansatz}, of the even spinor $\lambda_+$, while the reflected current comprises both the even $\lambda_+$ and the odd $\rho_-$ with different sign.

In the strongly interacting half-space, the Mott-N\'eel structure fixes two components to zero. Together with $N_\mu^{0A}=N_\mu^{1B}=1$ there is only $p_\mu^{0A} \neq 0$ and $p_\mu^{1B} \neq 0$. Therefore the current reads 
\begin{equation}
	J_\mu^\mathrm{Mott} =
	\begin{cases}
		i\frac{T}{Z} J_\mu^\mathrm{even} & \text{inside Mott bands,} \\
		i\frac{T}{Z}J_\mu^\mathrm{odd} & 0<E<U, \\
		0 & E<0 \text{ or } E>U
	\end{cases}
\end{equation}
with 
\begin{equation}
\begin{aligned}
	2J_\mu^\mathrm{even}=&\left( \tilde{A}_i \tilde{B}_i^*+ \tilde{A}_i^* \tilde{B}_i\right)\left(\kappa_i - \kappa_i^{-1} \right) \\&+\left( \tilde{A}_j \tilde{B}_j^*+ \tilde{A}_j^* \tilde{B}_j\right)\left(\kappa_j - \kappa_j^{-1} \right), \\
    2J_\mu^\mathrm{odd}=& \kappa_i^{2}\kappa_j^{-1} \left(\tilde{B}_i \tilde{A}_j^*+ \tilde{A}_i \tilde{B}_j^* \right) \\&- \kappa_i^{-2}\kappa_j^1 \left(\tilde{A}_j \tilde{B}_i^*+\tilde{B}_j \tilde{A}_i^* \right) \\
		&+ \kappa_j^{2} \kappa_i^{-1} \left(\tilde{B}_j  \tilde{A}_i^*+\tilde{A}_j \tilde{B}_i^*\right)\\&- \kappa_j^{-2} \kappa_i^1 \left(\tilde{A}_i \tilde{B}_j^*+\tilde{B}_i \tilde{A}_j^* \right).
    \end{aligned}
\end{equation}

\begin{figure}
	\centering
	 					\tikzsetnextfilename{egreaterjplotmottneel}
	\begin{tikzpicture}
		\begin{axis}[
			xlabel={$E/U$},
			ylabel={$J^\mathrm{in}$},
			ymin=0,
			xtick distance=0.05,
			legend style={at={(0.5,1.1)}, anchor=north, legend columns=5},
				minor tick num = 1,
		major grid style = {lightgray},
		minor grid style = {lightgray!25},
			width=1\linewidth, 
			height=0.7\linewidth 
			]
			\addplot[smooth, no markers, teal] table[x index=0, y index=1] {pics/ref.txt};
			\addlegendentry{$-J^\mathrm{ref}$};
			
			\addplot[smooth, no markers,blue] table[x index=0, y index=1] {pics/refsym.txt};
			\addlegendentry{$-J^\mathrm{ref,sym}$};
			
			\addplot[smooth, no markers,red] table[x index=0, y index=1] {pics/refasum.txt};
			\addlegendentry{$-J^\mathrm{ref,asym}$};
			
			\addplot[smooth, no markers,orange] table[x index=0, y index=1] {pics/transsym.txt};
			\addlegendentry{$J^\mathrm{even}$};
			
			\addplot[smooth, no markers,black] table[x index=0, y index=1] {pics/transasym.txt};
			\addlegendentry{$J^\mathrm{odd}$};

            \draw[dashed] (1,0) -- (1,1.5);
            \draw[dashed] (1.025,0) -- (1.025,1.5);
		\end{axis}
	\end{tikzpicture}
\caption[Currents in single interface]{Reflected and transmitted currents as function of energy in a two-dimensional lattice with $V= 1.1U$, $T= 0.2U$, $Z=4$ and $k^\|=0.3 \pi$ for energies greater than half the band gap, $E>U/2$. Dashed vertical lines mark the edges of the upper Hubbard band.}
\label{fig:egruecurrents}
\end{figure}
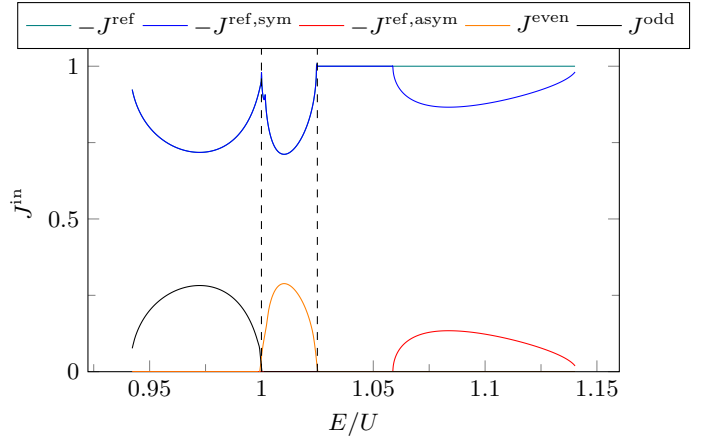
In these equations, we introduced the abbreviations $\tilde{A}_i=\mathcal{A}_i/N_i$ and $\tilde{B}_i=\mathcal{B}_i/N_i$.
These quasi-particle currents are independent of the lattice site $\mu$. From the currents we can already read off that perfect reflection only happens outside the Mott bands for $E>U$ and $E<0$. Inside the Mott bands, there is transmission of the even current contribution, inside the gap of the odd one. 
For any parameter combination, independent of the energy, the conservation of the quasi-particle current 
\begin{equation}
	J^\mathrm{even} +J^\mathrm{odd} - J^\mathrm{ref}= J^\mathrm{in}
\end{equation}
always holds. The negative sign in front of the reflected current is due to the different propagation direction.

Because of the sign change in the group velocity in the middle of the band gap at $E=U/2$, we need to distinguish these two cases: 
For energies above $U/2$ -- where the transmission is dominated by doublons -- and  below -- where transmission is dominated by holons -- we find different physics. 

Furthermore, we will distinguish between reflection from the incoming even spinor into the even spinor, denoted as $J^{\mathrm{ref,sym}}$, and from the even into the odd one, denoted as $J^{\mathrm{ref,asym}}$. Their sum is the total reflected current $J^\mathrm{ref}=J^{\mathrm{ref,sym}}+J^{\mathrm{ref,asym}}$. With this, we can also define the transmission $T=J^\mathrm{Mott}/J^\mathrm{in}$ and reflection coefficient $R=-J^\mathrm{ref}/J^\mathrm{in}$, respectively. The additional minus sign compensates the different propagating directions.

\subsection{Regular Reflection}
\blue{As a first example, we study the scattering of a particle hitting the interface from the weakly interacting half-space. For $E > U/2$, the reflected and transmitted current are plotted in  
Fig.~\ref{fig:egruecurrents}. We discuss the reflection as function of the energy $E$.}
As already expected from the band overlap in Fig.~ \ref{fig:bandersemioverlap}, there is reflection from the incoming even into the odd spinor (red line). It sets on as soon as the bands energetically overlap, but remains smaller than the reflection back into the even spinor (blue line). 
Inside the band gap there is transmission via the odd current combination (black line), whereas inside the upper Hubbard band the transmission happens via the even one \blue{alone} (orange line). For all energies $0 \leq T \leq 1$ and $0 \leq R \leq 1$ holds. \blue{In summary, we observe} the usual transmission characteristics at interfaces already known from standard quantum mechanics problems. 

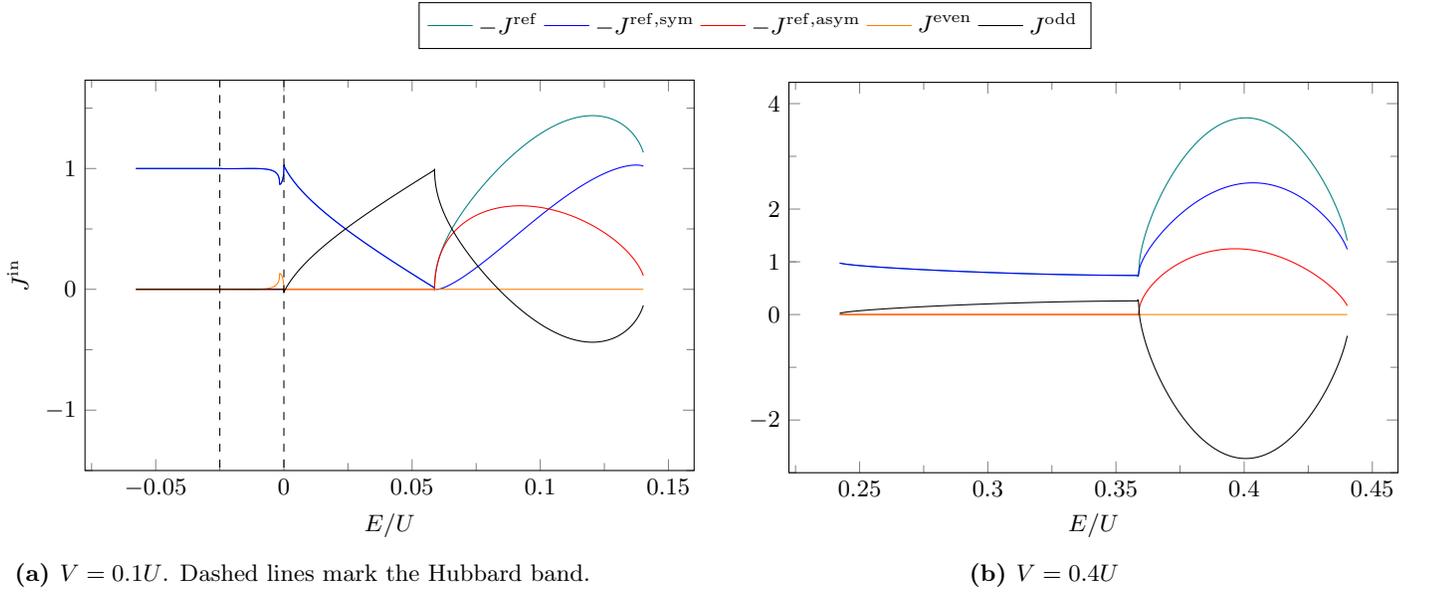
\begin{figure*}
	\centering
	\begin{subfigure}{0.45\textwidth}
		\centering
			\tikzsetnextfilename{esmallerjplotmottneel}
		\begin{tikzpicture}
		\begin{axis}[
			xlabel={$E/U$},
			ylabel={$J^\mathrm{in}$},
			ymin=-1.5,
			xtick distance=0.05,
			legend style={at={(1.1,1.2)}, anchor=north, legend columns=6},
				minor tick num = 1,
		major grid style = {lightgray},
		minor grid style = {lightgray!25},
		width = 1.2\textwidth,
		height = 0.84\linewidth,
		 scaled ticks=false, 
		xticklabel style={/pgf/number format/fixed}
			]
			\addplot[smooth, no markers, teal] table[x index=0, y index=1] {pics/2ref.txt};
			\addlegendentry{$-J^\mathrm{ref}$};
			
			\addplot[smooth, no markers,blue] table[x index=0, y index=1] {pics/2refsym.txt};
			\addlegendentry{$-J^\mathrm{ref,sym}$};
			
			\addplot[smooth, no markers,red] table[x index=0, y index=1] {pics/2refasum.txt};
			\addlegendentry{$-J^\mathrm{ref,asym}$};
			
			\addplot[smooth, no markers,orange] table[x index=0, y index=1] {pics/2transsym.txt};
			\addlegendentry{$J^\mathrm{even}$};
			
			\addplot[smooth, no markers,black] table[x index=0, y index=1] {pics/2transasym.txt};
			\addlegendentry{$J^\mathrm{odd}$};
			
               \draw[dashed] (0,-2) -- (0,2.5);
            \draw[dashed] (-.025,-2) -- (-0.025,2.5);
		\end{axis}
	\end{tikzpicture}
	\caption[Currents in single interface]{$V= 0.1U$. Dashed lines mark the Hubbard band.}
		\label{fig:sub1}
	\end{subfigure}
	\hfill
	\begin{subfigure}{0.45\textwidth}
	\centering
	\tikzsetnextfilename{esmallerjplotmottneel3}
	\begin{tikzpicture}
		\begin{axis}[
			xlabel={$E/U$},
			ymin=-3,
			xtick distance=0.05,
			legend style={at={(0.5,1)}, anchor=north, legend columns=6},
				minor tick num = 1,
		major grid style = {lightgray},
		minor grid style = {lightgray!25},
			width = 1.2\textwidth,
			height = 0.84\linewidth,
			extra y ticks={1}
			]
			\addplot[smooth, no markers, teal] table[x index=0, y index=1] {pics/3ref.txt};
			
			\addplot[smooth, no markers,blue] table[x index=0, y index=1] {pics/3refsym.txt};
			
			\addplot[smooth, no markers,red] table[x index=0, y index=1] {pics/3refasum.txt};
			
			\addplot[smooth, no markers,orange] table[x index=0, y index=1] {pics/3transsym.txt};
			
			\addplot[smooth, no markers,black] table[x index=0, y index=1] {pics/3transasym.txt};
			
		\end{axis}
	\end{tikzpicture}
	\caption[Currents in single interface]{$V= 0.4U$}
	\label{fig:sub2}
\end{subfigure}
	\caption[Currents in single interface]{Reflected and transmitted currents in units of the incoming one as function of energy in a two-dimensional lattice with $T= 0.2U$, $Z=4$ and $k^\|=0.3 \pi$ for energies smaller than half the band gap, $E<U/2$.}
	\label{fig:main}
\end{figure*}

\subsection{Analogy to the Klein Paradox}

\blue{Different behavior is encountered for $E<U/2$, as }
shown in Fig.~\ref{fig:main} for two choices of the on-site potential. For such energies the transmission in the Mott insulator is holon-dominated.

Generally,  inside the lower Hubbard band the transmission is smaller compared to the upper one and strongly depends on the hopping strength. Larger hopping increases the transmission probability. 

There is still reflection into both spinors, into the even (blue line) and odd (red line) one. 
Different from the previous case $E>U/2$, now $R>1$ becomes possible 
inside the band gap. This is 
always compensated by $T<0$ (black line), such that $R+T=1$, to conserve the total current. 
Subject to an incoming electron, 
stimulated emission of doublon-holon pairs occurs at the interface  with an amplitude given by $|T(E)|$ \cite{wagner2010bosonic}.

As the energy $E$ is increased, there is always an initial increase of 
$J^\mathrm{odd}$ prior to the sign change of $T$ from positive to negative.  
This is similar to the Klein paradox in graphene, that shows perfect transmission because of the chirality of the particles \cite{he2013chiral,katsnelson2006chiral}, or the one for magnons at ferromagnetic interfaces, where the bosonic Klein paradox analog leads to a stimulated emission and enhances spin currents \cite{Kleinherbers2025,Bassant2024,Harms2022}.

The amplitude of the Klein paradox increases with equalizing the amount of doublons and holons taking part in the transmission. This increases from purely holon dominated in the lower Hubbard band towards a ratio of one in the middle of the band gap. The closer the energy is to this midpoint, the larger the amplitude, as shown in Fig.~\ref{fig:sub1} and Fig.~\ref{fig:sub2}, such that the reflection from the even into the even  $J^\mathrm{ref,sym}$ and into the odd $J^\mathrm{ref,asym}$ exceeds one. For the maximum amplitude, the symmetric reflection is always greater than the antisymmetric one. The two amplitudes are shown in Fig.~\ref{fig:maindens} as a function of the parallel momentum and energy as a color plot. $J^\mathrm{ref,sym}$ shows the Klein paradox analog for small energies and small parallel momentum, while $J^\mathrm{ref,asym}$ shows it for larger energies and large parallel momentum.

\subsection{\red{Effective Dirac Equation}}
In order to understand {the origin of the Klein paradox analog in this condensed matter system we will derive an effective Dirac equation governing the dynamics in a long wavelength limit.

Starting in real space, the two coupled equations read:
\begin{equation}
	(	i \partial_t-U^I)p_\mu^{I_{X}}=-\sum_{\kappa} \frac{T_{\mu \kappa}}{Z} p_{\kappa}^{\bar{I}_{\bar{X}}}.
\label{eq:coupled}
\end{equation}
\blue{Here, the notation $\bar I$ and $\bar X$ denotes the opposite value of $I$ and $X$, respectively.} 
Only hole excitations $I=0$ of spin $\uparrow$ and particle excitations $I=1$ with spin $\downarrow$ are supported on sublattice $A$, and vice versa on $B$. All other ones are trivial and omitted. Therefore, particle excitations on $A$ are tunnel-coupled to hole excitations on $B$ and vice versa.
We introduce the effective spinor 
\begin{equation}
    \psi_{\mu\in A} = \begin{pmatrix}\h c_{\mu \downarrow I=1} \\ \h c_{\mu \uparrow I=0} \end{pmatrix} \quad \psi_{\mu\in B} = \begin{pmatrix}\h c_{\mu \uparrow I=1} \\ \h c_{\mu \downarrow I=0} \end{pmatrix},
\end{equation}
which allows us to express the coupled system \ref{eq:coupled} as:
\begin{equation}
	i \partial_t \psi_\mu =
	\begin{pmatrix}
		U & 0 \\ 0 & 0
	\end{pmatrix}\psi_\mu
	-\frac{1}{Z}\sum_\kappa T_{\mu \kappa}\sigma_x \psi_\kappa .
\end{equation}
Applying the trivial phase redefinition $\psi_\mu \to e^{-i U t/2}\psi_\mu$, this becomes:
\begin{equation}
	i \partial_t \psi_\mu = \frac{U}{2}\sigma_z \psi_\mu
	-\frac{1}{Z}\sum_\kappa T_{\mu \kappa}\sigma_x \psi_\kappa
\end{equation}
with the Pauli matrices $\sigma_x$ and $\sigma_z$.
At this point we aim to take the continuum limit, which is appropriate when the quasi-particle wavelength is much larger than the lattice spacing.  
As a first step, we derive the dispersion relation. Moving to Fourier space we go from the real space hopping element $T_{\mu \nu}$ to the momentum dependent coefficient $T_\mathbf{k}$ and find:
\begin{equation}
\omega(\mathbf{k}) \psi_\mathbf{k}
= \Bigl(\tfrac{U}{2}\sigma_z - T_\mathbf{k}\sigma_x\Bigr)\psi_\mathbf{k},
\end{equation}
which leads to:
\begin{equation}
	\omega(\mathbf{k}) = \pm \sqrt{\tfrac{U^2}{4} + T_\mathbf{k}^2}.
\end{equation}
The minimum \blue{gap} of the spectrum occurs when $T_{\mathbf{k}_0}=0$ (more precisely, $T_{\mathbf{k}_0}^2=0$).  
Expanding $T_\mathbf{k}$ around this zero momentum, $\mathbf{k}=\mathbf{k}_0 + \mathbf{q}$ yields:
\begin{equation}
	T_\mathbf{k}\,\sigma_x \psi_\mathbf{k}
	= \mathbf{q}\cdot
	\Bigl(\nabla_\mathbf{k} T_\mathbf{k}\big|_{\mathbf{k}_0}\Bigr)
	\sigma_x \psi_\mathbf{k}
	= \mathbf{q}\cdot \mathbf{c}_\mathrm{eff}\,\sigma_x \psi_\mathbf{k},
\end{equation}
where we identify the effective propagation velocity as:
$\mathbf{c}_\mathrm{eff}=\nabla_\mathbf{k} T_\mathbf{k}\big|_{\mathbf{k}_0}$.

\begin{figure*}
	\centering
	\begin{subfigure}{0.45\textwidth}
		\centering
		\tikzsetnextfilename{densityodd}
		\begin{tikzpicture} 
			
			\begin{axis}[xshift=8cm,
				xlabel=$k^\|/\pi$,
				ylabel=$E/U$,
				small,view={0}{90},colorbar,
						minor tick num = 1,
				major grid style = {lightgray},
				minor grid style = {lightgray!25},
				yticklabel style={/pgf/number format/fixed, /pgf/number format/precision=2}, 
				point meta min=0, 
				point meta max=4.5,  
				]
				\addplot3 [surf,
				shader=flat,
				]
				table {pics/outputodd.txt};
			\end{axis}
		\end{tikzpicture}
		\caption[Currents in single interface]{$-J^\mathrm{refl,asym}$}
		\label{fig:sub1odd}
	\end{subfigure}
	\quad \quad
	\begin{subfigure}{0.45\textwidth}
		\centering
		\tikzsetnextfilename{densityeven}
		\begin{tikzpicture} 
			\begin{axis}[
				xlabel=$k^\|/\pi$,
				yticklabel pos=right,
					ylabel=$E/U$,
				small,view={0}{90},
						minor tick num = 1,
				major grid style = {lightgray},
				minor grid style = {lightgray!25},
				yticklabel style={/pgf/number format/fixed, /pgf/number format/precision=2} ,
				point meta min=0, 
				point meta max=4.5,  
				]
				\addplot3 [surf,
				shader=flat,
				]
				table {pics/outputeven.txt};
			\end{axis}
		\end{tikzpicture}
		\caption[Currents in single interface]{$-J^\mathrm{refl,sym}$}
		\label{fig:sub2even}
	\end{subfigure}
	\caption[Currents in single interface]{Density plot of the reflected current for the a) asymmetric and b) symmetric reflection as a function of parallel momentum and energy. The plots use the same color scaling, the parameters are $V=0.2U$, $T=0.2U$ in a two-dimensional lattice.}
	\label{fig:maindens}
\end{figure*}
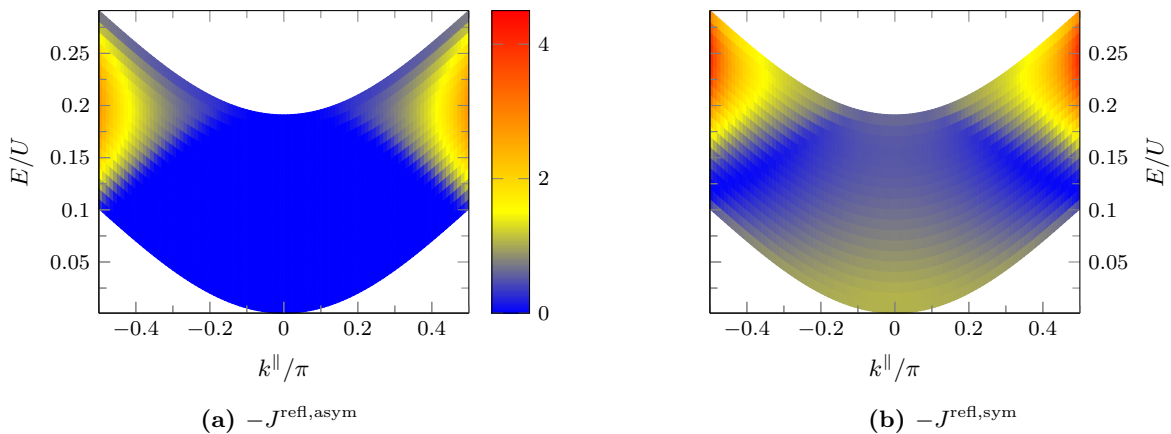

Substituting this back, the equation becomes:
\begin{equation}
	i \partial_t \psi_\mathbf{k} =
	\left( \tfrac{U}{2}\sigma_z
	- \mathbf{c}_\mathrm{eff}\cdot \mathbf{q}\,\sigma_x \right)\psi_\mathbf{k}.
\end{equation}
Upon replacing $\mathbf{q}\to -i\nabla$ and Fourier transforming to real space, we arrive at the Dirac equation 
\blue{in $\texttt{1+1}$ dimensions}
with a finite mass term $U$, separating the upper and lower Hubbard bands.

A parallel calculation for a bipartite semiconductor background leads instead to a Weyl-type equation: the eigenmodes of the two sublattices become coupled, but with zero mass gap. This occurs at momentum $\pi/2$, where both the weakly and strongly correlated layers reach their minimum energy gap.
\subsection{Angle-of-incidence Dependence}
\label{subsec:angledependent}
The incident angle $\varphi$ of the particles can be calculated from the dispersion relation Eq.~\ref{semiconductor} and the parallel momentum $k^\|$ via 
\begin{equation}
\begin{aligned}
	\varphi &= \arctan\left(\frac{k^\|}{\kappa_{\mathrm{semi}}}\right)\\&=\arctan\left(\frac{k^\|}{\arccos\left[\frac{Z}{2T}(V-E)-\cos k^\|\right]}\right).
    \end{aligned}
\end{equation}
Note that this equation does not yield a solution for any arbitrary triples of 
energy $E$, on-site potential $V$, hopping strength $T$ and parallel momentum $\kappa_{\mathrm{semi}}$.
The angle dependent reflection $R=-\frac{J^\mathrm{ref}}{J^\mathrm{in}}$  is shown in Fig.~\ref{fig:polarplots} for three different band alignments. Fig.~\ref{fig:sub0polar} shows the $E>U/2$ case, Fig.~\ref{fig:sub1polar} and Fig.~\ref{fig:sub2polar} the $E<U/2$ case for $V=0.4U$ and $V=0.2U$, respectively.

For $E>U/2$, the minimal reflection 
is found for energies inside the upper Hubbard band (e.g. $E=1.01U$) for $\varphi=0$ and transmission decreases with increasing angle, while for energies below the Hubbard band edge 
(e.g. $E=0.96U$ and $E=0.99U$) 
minima are found for $\varphi \neq 0$. The relation is reversed, increasing the angle increases transmission, until reflection from the even into the odd spinor turns on, see the kinks in Fig. ~\ref{fig:sub00polar}. This shows $-J^\mathrm{ref}$, $-J^\mathrm{ref,sym}$ and $-J^\mathrm{ref,asym}$ for an energy slightly below the upper Hubbard band.

The analogous behavior 
in the Klein paradox regime, $E<U/2$, can be seen in Fig.~\ref{fig:sub1polar} and Fig.~\ref{fig:sub2polar}: if the energy for a certain band alignment does not allow for the reflection to be larger than one, it decreases with increasing the angle until reflection into the odd spinor turns on (blue line in Fig.~\ref{fig:sub2polar}). If above-unity 
reflection is possible, its magnitude increases with increasing the angle.
\begin{figure*}
	\begin{subfigure}{0.45\textwidth}
	\centering
		\tikzsetnextfilename{polarplotthree}
\begin{tikzpicture}
	\begin{polaraxis}[
		grid=both,
		xlabel={angle $\varphi$},
		   ylabel={$R$},
		ytick={0.25,0.5,0.75},
		extra y ticks={1},
		extra y tick style={
			color=red, 
			grid style={color=red} 
		},
		xticklabel style={sloped like x axis},
		xtick={270,315,360,405,450},
		xticklabels={$-\pi/2$,$-\pi/4$,0,$\pi/4$,$\pi/2$},
		xmin=270,xmax=450,
		ymax=1.1,
		legend style={at={(1,0.6)},anchor=south west, legend cell align=right},
		yticklabel pos=left,
		yticklabel style={
			anchor=east,
		},
		]
		\addplot[blue, thick, data cs=polarrad] table[col sep=space] {pics/angletable_u_1_v_1.05_t_0.2_z_4_e_0.96.dat};
		\addplot[green, thick, data cs=polarrad] table[col sep=space] {pics/angletable_u_1_v_1.05_t_0.2_z_4_e_0.99.dat};
		\addplot[orange, thick, data cs=polarrad] table[col sep=space] {pics/angletable_u_1_v_1.05_t_0.2_z_4_e_1.01.dat};
		
		\legend{$E=0.96U$,$0.99U$,$1.01U$}

	\end{polaraxis}
\end{tikzpicture}

	\caption{Reflection $R$ for $V= 1.05U$}
	\label{fig:sub0polar}
\end{subfigure}
	\begin{subfigure}{0.45\textwidth}
	\centering
	\tikzsetnextfilename{polarplotfour}
\begin{tikzpicture}
	\begin{polaraxis}[
		grid=both,
			xlabel={angle $\varphi$},
		ylabel={$j^\mathrm{in}$},
		ytick={0.25,0.5,0.75},
		extra y ticks={1},
		extra y tick style={
			color=red, 
			grid style={color=red} 
		},
		xticklabel style={sloped like x axis},
		xtick={270,315,360,405,450},
		xticklabels={$-\pi/2$,$-\pi/4$,0,$\pi/4$,$\pi/2$},
		xmin=270,xmax=450,
		ymax=1.1,
		legend style={at={(1,0.6)},anchor=south west, legend cell align=left},
		yticklabel pos=left,	yticklabel style={
			anchor=east,
		},
		]
		\addplot[blue, thick, data cs=polarrad] table[col sep=space] {pics/angletable_u_1_v_1.05_t_0.2_z_4_e_0.99.dat};
		\addplot[green, thick, data cs=polarrad] table[col sep=space] {pics/evenangletable_u_1_v_1.05_t_0.2_z_4_e_0.99.dat};
		\addplot[orange, thick, data cs=polarrad] table[col sep=space] {pics/oddangletable_u_1_v_1.05_t_0.2_z_4_e_0.99.dat};
		
		\legend{$-J^\mathrm{ref}$,$-J^\mathrm{ref,sym}$,$-J^\mathrm{ref,asym}$}

	\end{polaraxis}
\end{tikzpicture}
	
	\caption{$V= 1.05U$, $E=0.99U$}
	\label{fig:sub00polar}
\end{subfigure}

\begin{subfigure}{0.45\textwidth}
	\tikzsetnextfilename{polarplotone}
\begin{tikzpicture}
	\begin{polaraxis}[
		grid=both,
			xlabel={angle $\varphi$},
		ylabel={$R$},
		ytick={5,10},
		extra y ticks={1},
		extra y tick style={
			color=red, 
			grid style={color=red} 
		},
		xticklabel style={sloped like x axis},
		xtick={270,315,360,405,450},
		xticklabels={$-\pi/2$,$-\pi/4$,0,$\pi/4$,$\pi/2$},
		xmin=270,xmax=450,
		ymax=20,
		legend style={at={(1,0.6)},anchor=south west, legend cell align=right},
		yticklabel pos=left,	yticklabel style={
			anchor=east,
		},
		]
		\addplot[blue, thick, data cs=polarrad] table[col sep=space] {pics/angletable_u_1_v_0.4_t_0.2_z_4_e_0.3.dat};
		\addplot[green, thick, data cs=polarrad] table[col sep=space] {pics/angletable_u_1_v_0.4_t_0.2_z_4_e_0.35.dat};
		\addplot[orange, thick, data cs=polarrad] table[col sep=space] {pics/angletable_u_1_v_0.4_t_0.2_z_4_e_0.4.dat};
		
		\legend{$E=0.3U$,$0.35U$,$0.4U$}

	\end{polaraxis}
\end{tikzpicture}
	\caption{Reflection $R$ for $V= 0.4U$}
	\label{fig:sub1polar}
\end{subfigure}
	\begin{subfigure}{0.45\textwidth}
	\centering
	\tikzsetnextfilename{polarplottwo}
\begin{tikzpicture}
	\begin{polaraxis}[
		grid=both,
			xlabel={angle $\varphi$},
		ylabel={$R$},
		ytick={0.5,1.5},
		extra y ticks={1},
		extra y tick style={
			color=red, 
			grid style={color=red} 
		},
		xticklabel style={sloped like x axis},
		xtick={270,315,360,405,450},
		xticklabels={$-\pi/2$,$-\pi/4$,0,$\pi/4$,$\pi/2$},
		xmin=270,xmax=450,
		ymax=2,
		legend style={at={(1,0.6)},anchor=south west, legend cell align=right},
		yticklabel pos=left,	yticklabel style={
			anchor=east,
		},
		]
		\addplot[blue, thick, data cs=polarrad] table[col sep=space] {pics/angletable_u_1_v_0.2_t_0.2_z_4_e_0.12.dat};
		\addplot[orange, thick, data cs=polarrad] table[col sep=space] {pics/angletable_u_1_v_0.2_t_0.2_z_4_e_0.14.dat};
		
		\legend{$E=0.12U$,$0.14U$}

	\end{polaraxis}
\end{tikzpicture}
	\caption{Reflection $R$ for $V= 0.2U$}
	\label{fig:sub2polar}
\end{subfigure}
\caption[Angle dependence in single interface]{Incident angle dependence of the reflection $R$ and $-J^\mathrm{ref,sym}$, $-J^\mathrm{ref,asym}$for different on-site potentials and energies. The radial axis is cropped for better visibility.}
\label{fig:polarplots}
\end{figure*}
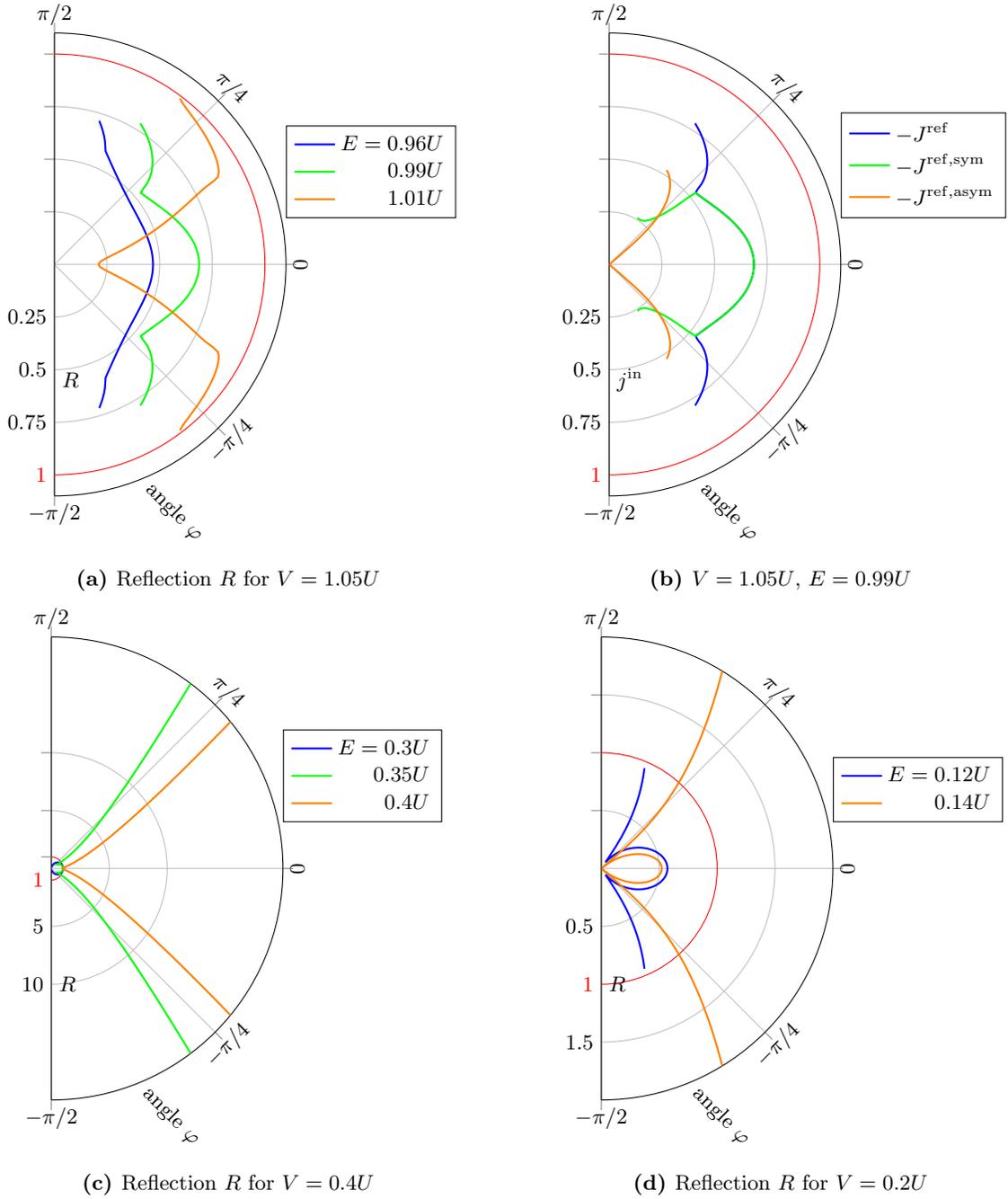

\section{Conclusions}
Using the hierarchy of correlations, we investigated a single interface between a weakly interacting layer and a strongly interacting Mott insulator \red{in the Mott-Neel 
type spin ordered background. We found an analogy to quantum electrodynamics, as in the vicinity of the minimum gap the doublons and holons are effectively described by a Dirac equation}. We derived the propagation of the electrons as well as the doublons and holons in their respective regions. For an incoming electron, we calculated the reflection and transmission characteristics, showing that for energies 
exceeding half the band gap the system behaves 
like a potential barrier known from standard quantum mechanics.
For energies below, we found behavior 
reminiscent of the Klein paradox, with reflections greater than one and negative transmission coefficients. This shows that this interface is another condensed matter system 
that can serve as an analog to the Klein paradox. \red{We note that the creation of many doublon-holon pairs in the Klein paradox analog would alter the mean-field background $\h \rho^0$ significantly, such that the back-reaction of this onto the mean-field background, an effect included in the higher-order equation $i \partial_t \h \rho_\mu= F_1(\hat{\rho}_\mu, \hat{\rho}^{\text{corr}}_{\mu \nu})$, should be incorporated. Additionally, the heating due to the creation of many pairs might destroy the spin order necessary for the analogy to QED. Therefore, our analysis only describes the onset of the dielectric breakdown of the Mott insulator. This is the subject of future works.}

\section{Acknowledgment}
The authors thank F. Queisser and R. Schützhold for fruitful discussions and valuable feedback on the manuscript. This work is funded by the Deutsche Forschungsgemeinschaft (DFG, German Research Foundation) – Project-ID 278162697– SFB 1242.
\appendix

\section{Hierarchy of Correlations}
\label{app:hierarchyofcorrelations}
In order to describe the (quasi-)particles in the heterostructure, we employ the \emph{hierarchy of correlations} approach~\cite{Queisser2014,Queisser2019,PhysRevA.82.063603}. This method starts from a general lattice Hamiltonian of the form:
\begin{equation}
    \label{eq:generalManyBodyHam}
    \hat{H} = \frac{1}{Z} \sum_{\mu \nu} \hat{H}_{\mu \nu} + \sum_\mu \hat{H}_\mu,
\end{equation}
where $\mu$ and $\nu$ are generalized lattice coordinates and $Z$ is the coordination number. In the regime of large coordination number $Z \gg 1$, a controlled truncation scheme becomes possible, yielding a closed and iterative set of equations.

The approach begins with the Heisenberg equation of motion for the density operator $\hat{\rho}$:
\begin{equation}
    i \partial_t \hat{\rho} = [\hat{H}, \hat{\rho}] = \frac{1}{Z} \sum_{\mu \nu} \widehat{\mathcal{L}}_{\mu \nu} \hat{\rho} + \sum_\mu \widehat{\mathcal{L}}_\mu \hat{\rho},
\end{equation}
where the Liouville superoperators are defined as $\widehat{\mathcal{L}}_{\mu \nu}(\hat{\rho}) = [\hat{H}_{\mu \nu}, \hat{\rho}]$ and $\widehat{\mathcal{L}}_\mu(\hat{\rho}) = [\hat{H}_\mu, \hat{\rho}]$.

Since physical observables are typically defined over a subset of lattice sites, we decompose the reduced density matrices into uncorrelated and correlated parts:
\begin{equation}
    \begin{aligned}
        \hat{\rho}_{\mu \nu} &= \hat{\rho}^{\text{corr}}_{\mu \nu} + \hat{\rho}_\mu \hat{\rho}_\nu, \\
        \hat{\rho}_{\mu \nu \lambda} &= \hat{\rho}^{\text{corr}}_{\mu \nu \lambda}
        + \hat{\rho}^{\text{corr}}_{\mu \nu} \hat{\rho}_\lambda
        + \hat{\rho}^{\text{corr}}_{\mu \lambda} \hat{\rho}_\nu
        + \hat{\rho}^{\text{corr}}_{\nu \lambda} \hat{\rho}_\mu
        + \hat{\rho}_\mu \hat{\rho}_\nu \hat{\rho}_\lambda,
    \end{aligned}
    \label{eq:splittingcorr}
\end{equation}
and so on for higher-order correlations.

The time evolution of the on-site density operator $\hat{\rho}_\mu$ is given by:
\begin{equation}
    i \partial_t \hat{\rho}_\mu = \frac{1}{Z} \sum_{\alpha \neq \mu} \mathrm{tr}_\alpha \left( \widehat{\mathcal{L}}^S_{\alpha \mu} \left[ \hat{\rho}^{\text{corr}}_{\mu \alpha} + \hat{\rho}_\alpha \hat{\rho}_\mu \right] \right) + \widehat{\mathcal{L}}_\mu \hat{\rho}_\mu,
\end{equation}
where the symmetrized Liouvillian is $\widehat{\mathcal{L}}^S_{\mu \nu} = \widehat{\mathcal{L}}_{\mu \nu} + \widehat{\mathcal{L}}_{\nu \mu}$. This equation couples to the two-point correlator $\hat{\rho}^{\text{corr}}_{\mu \nu}$.

A similar procedure applies to the evolution of $\hat{\rho}^{\text{corr}}_{\mu \nu}$, resulting in:
\begin{equation}
    \begin{aligned}
        i \partial_t \hat{\rho}^{\text{corr}}_{\mu \nu} =&\, \widehat{\mathcal{L}}_\mu \hat{\rho}^{\text{corr}}_{\mu \nu}
        + \frac{1}{Z} \widehat{\mathcal{L}}_{\mu \nu} \left( \hat{\rho}^{\text{corr}}_{\mu \nu} + \hat{\rho}_\mu \hat{\rho}_\nu \right) \\
        & - \frac{\hat{\rho}_\mu}{Z} \, \mathrm{tr}_\mu \left( \widehat{\mathcal{L}}^S_{\mu \nu} \left[ \hat{\rho}^{\text{corr}}_{\mu \nu} + \hat{\rho}_\mu \hat{\rho}_\nu \right] \right) \\
        & + \frac{1}{Z} \sum_{\alpha \neq \mu, \nu} \mathrm{tr}_\alpha \left( \widehat{\mathcal{L}}^S_{\mu \alpha} \left[ \hat{\rho}^{\text{corr}}_{\mu \nu \alpha}
        + \hat{\rho}^{\text{corr}}_{\mu \nu} \hat{\rho}_\alpha
        + \hat{\rho}^{\text{corr}}_{\nu \alpha} \hat{\rho}_\mu \right] \right) \\
        & + (\mu \leftrightarrow \nu).
    \end{aligned}
    \label{time_evo_rho_munucorr}
\end{equation}
This equation introduces the three-point correlator $\hat{\rho}^{\text{corr}}_{\mu \nu \lambda}$, and so the equations form an infinite hierarchy:
\begin{eqnarray}
    i \partial_t \hat{\rho}_\mu &=& F_1(\hat{\rho}_\mu, \hat{\rho}^{\text{corr}}_{\mu \nu}), \nonumber \\
    i \partial_t \hat{\rho}^{\text{corr}}_{\mu \nu} &=& F_2(\hat{\rho}_\mu, \hat{\rho}^{\text{corr}}_{\mu \nu}, \hat{\rho}^{\text{corr}}_{\mu \nu \lambda}), \nonumber \\
    i \partial_t \hat{\rho}^{\text{corr}}_{\mu \nu \lambda} &=& F_3(\hat{\rho}_\mu, \hat{\rho}^{\text{corr}}_{\mu \nu}, \hat{\rho}^{\text{corr}}_{\mu \nu \lambda}, \hat{\rho}^{\text{corr}}_{\mu \nu \lambda \kappa}), \nonumber \\
    i \partial_t \hat{\rho}^{\text{corr}}_{\mu \nu \lambda \alpha} &=& F_4(\hat{\rho}_\mu, \hat{\rho}^{\text{corr}}_{\mu \nu}, \hat{\rho}^{\text{corr}}_{\mu \nu \lambda}, \hat{\rho}^{\text{corr}}_{\mu \nu \lambda \kappa}, \hat{\rho}^{\text{corr}}_{\mu \nu \lambda \kappa \beta}).
    \label{evolution_appendix}
\end{eqnarray}
The exact form of the functionals $F_n$ depends on the specific structure of the Hamiltonian.

If the initial state exhibits a scaling behavior where $\ell$-point correlations scale as $\mathcal{O}(Z^{-\ell+1})$, this property is preserved throughout time evolution~\cite{Queisser2014,queisser2023hierarchy}. This scaling allows for a systematic truncation of the hierarchy. Keeping only the leading orders yields:
\begin{equation}
    \begin{aligned}
        i \partial_t \hat{\rho}_\mu &\approx F_1(\hat{\rho}_\mu, 0), \quad \text{with solution } \hat{\rho}_\mu^{0}, \\
        i \partial_t \hat{\rho}^{\text{corr}}_{\mu \nu} &\approx F_2(\hat{\rho}_\mu^{0}, \hat{\rho}^{\text{corr}}_{\mu \nu}, 0).
    \end{aligned}
\end{equation}
These two equations form the basis for describing the charge modes in the heterostructure. The quantity $\hat{\rho}_\mu^{0}$ captures the mean-field background, encoding the charge and spin structure of the system.

\section{Hierarchy for the Fermi Hubbard Model}
\label{app:hierarchyforFermiHubbard}
To apply the hierarchy of correlations to the Fermi-Hubbard model defined in Eq.~\ref{eq:FHMHamiltonian}, we begin by introducing the spin background. This is represented by the antiferromagnetic Mott-N\'eel state, in which the lattice is divided into two sublattices, $A$ and $B$, arranged in a checkerboard configuration:
\begin{equation}
	\hat \rho_\mu^0 = 
	\begin{dcases}
	\ket{\uparrow}_\mu\!\bra{\uparrow} & \mu \in A, \\
	\ket{\downarrow}_\mu\!\bra{\downarrow} & \mu \in B.
	\end{dcases}
\end{equation}

Irrespective of the choice of mean-field background, it is useful to define quasi-particle operators. These are inspired by the Hubbard $X$-operators~\cite{Hubbard1965,ovchinnikov2004hubbard} and composite operator approaches~\cite{mancini2004hubbard}, and are written as:
\begin{equation}
\hat c_{\mu s I} = \hat c_{\mu s} \hat n_{\mu\bar{s}}^I =
\begin{cases}
	\hat c_{\mu s}(1 - \hat n_{\mu\bar{s}}) & \text{for } I = 0, \\
	\hat c_{\mu s} \hat n_{\mu\bar{s}} & \text{for } I = 1,
\end{cases}
\end{equation}
where $I = 0$ and $I = 1$ label holons and doublons, respectively. These operators offer a more accurate description of the relevant physical processes, although they only approximate the actual quasi-particle creation and annihilation operators for holons and doublons~\cite{Avigo2020}. Here, $\bar{s}$ denotes the spin opposite to $s$.

We now consider the two-point correlation functions $\langle \hat c_{\mu s I}^\dagger \hat c_{\nu s J} \rangle$, whose time evolution is given by:
\begin{equation}
\begin{aligned}
i \partial_t \langle \hat c^\dagger_{\mu s I} \hat c_{\nu s J} \rangle^{\mathrm{corr}} &=
\frac{1}{Z} \sum_{\lambda L} T_{\mu\lambda}
\langle \hat n_{\mu\bar{s}}^I \rangle^0
\langle \hat c^\dagger_{\lambda s L} \hat c_{\nu s J} \rangle^{\mathrm{corr}} \\
&\quad - \frac{1}{Z} \sum_{\lambda L} T_{\nu\lambda}
\langle \hat n_{\nu\bar{s}}^J \rangle^0
\langle \hat c^\dagger_{\mu s I} \hat c_{\lambda s L} \rangle^{\mathrm{corr}} \\
&\quad + \left( U_\nu^J - U_\mu^I + V_\nu - V_\mu \right)
\langle \hat c^\dagger_{\mu s I} \hat c_{\nu s J} \rangle^{\mathrm{corr}} \\
&\quad + \frac{T_{\mu\nu}}{Z} \left(
\langle \hat n_{\mu\bar{s}}^I \rangle^0
\langle \hat n_{\nu s}^1 \hat n_{\nu\bar{s}}^J \rangle^0
- \langle \hat n_{\nu\bar{s}}^J \rangle^0
\langle \hat n_{\mu s}^1 \hat n_{\mu\bar{s}}^I \rangle^0
\right) \\&\quad+ \mathcal{O}(1/Z^2),
\end{aligned}
\end{equation}
where we define $U_\mu^I = I U_\mu$, so that $U_\mu^I = 0$ for $I = 0$ and $U_\mu^I = U_\mu$ for $I = 1$.

The dynamical evolution of these correlations can be further simplified using a factorization approach~\cite{navez2014quasi,Queisser2014}, allowing us to isolate the amplitudes for holons ($I,J=0$) and doublons ($I,J=1$).

In the bipartite lattice structure, each sublattice has a distinct spin occupation: sites on sublattice $X$ satisfy $\langle \hat n_{\mu_X \uparrow} \rangle = 0$ and $\langle \hat n_{\mu_X \downarrow} \rangle = 1$, while the opposite holds for sites on sublattice $Y$. The correlation functions can be expressed in terms of Fourier components with an additional index to distinguish sublattices:
\begin{equation}
	\langle \hat c^\dagger_{\mu_X s I} \hat c_{\nu_Y s J} \rangle^{\mathrm{corr}} = \left( p_{\mu s}^{I_X} \right)^* p_{\nu s}^{J_Y},
\end{equation}
where $X,Y \in \{A,B\}$ refer to the sublattices. These amplitudes can be conveniently grouped using a spinor notation, and their dynamics are governed by effective equations~\cite{verlage2024quasi}.

Assuming a highly symmetric (e.g., hypercubic) lattice permits a Fourier transform along the directions parallel to the interface:
\begin{equation}
	\begin{aligned}
		p_{\mu s}^{I_X} &= \frac{1}{\sqrt{N^\parallel}} \sum_{\mathbf{k}^\parallel} p_{n,\mathbf{k}^\parallel,s}^{I_X} e^{i \mathbf{k}^\parallel \cdot \mathbf{x}_\mu^\parallel}, \\
		T_{\mu\nu} &= \frac{Z}{N^\parallel} \sum_{\mathbf{k}^\parallel} T_{m,n,\mathbf{k}^\parallel} e^{i \mathbf{k}^\parallel \cdot \left( \mathbf{x}_\mu^\parallel - \mathbf{x}_\nu^\parallel \right)}.
	\end{aligned}
\end{equation}
For isotropic nearest-neighbor hopping with $T_n^\| = T_{n,n-1}^\perp = T$, the hopping matrix elements take the form:
\begin{equation}
\begin{aligned}
	T_{m,n,\mathbf{k}^\parallel} &= \frac{T_{\mathbf{k}^\parallel}^\parallel}{Z} \delta_{m,n}
	+ \frac{T}{Z} (\delta_{n,n-1} + \delta_{n,n+1}), \\
	T_{\mathbf{k}^\parallel}^\parallel &= 2T \sum_{x_i} \cos(p_{x_i}^\parallel) \equiv Z T_{\mathbf{k}}^\parallel,
\end{aligned}
\end{equation}
where $T_{\mathbf{k}}^\parallel$ captures the momentum-dependent in-plane hopping.

The resulting doublon and holon amplitudes obey the coupled equations:
\begin{equation}
\label{eq:mottneeldefeq}
\left( i \partial_t - U^j \right) p_{\mu s}^{I_X} = - \langle \hat n_{\mu_X}^I \rangle \sum_L \left[ T_{\mathbf{k}}^\parallel p_{\mu}^{L_{\bar{X}}} + \frac{T}{Z} \left( p_{\mu+1}^{L_{\bar{X}}} + p_{\mu-1}^{L_{\bar{X}}} \right) \right],
\end{equation}
where $\bar{X}$ denotes the sublattice opposite to $X$.

Finally, assuming without loss of generality that $\langle \hat n_{\mu_A \uparrow} \rangle = 0$ and $\langle \hat n_{\mu_B \downarrow} \rangle = 0$, it follows directly that
\[
E\, p_\mu^{1_A} = (E - U)\, p_\mu^{0_B} \equiv 0.
\]

\vfill
\bibliography{bib.bib}
\end{document}